\documentclass[twocolumn,superscriptaddress,
showpacs,preprintnumbers,amsmath,amssymb]{revtex4}

\usepackage{amssymb,amsmath,amsthm}
\usepackage{graphicx}

\theoremstyle{definition}

\newcommand{\bra}[1]{{\left\langle #1 \right|}}
\newcommand{\ket}[1]{{\left| #1 \right\rangle}}
\newcommand{\inn}[2]{{\left\langle #1 | #2 \right\rangle}}

\newcommand{\Z}{\mbox{$\mathbb Z$}}

\begin{document}
\title{Three-party $d$-level quantum secret sharing protocol}

\author{Dong~Pyo~Chi}\email{dpchi@math.snu.ac.kr}
\affiliation{
 Department of Mathematical Sciences,
 Seoul National University, Seoul 151-747, Korea
}
\author{Jeong~Woon~Choi}\email{cju@snu.ac.kr}
\affiliation{
 Department of Mathematical Sciences,
 Seoul National University, Seoul 151-747, Korea
}
\author{Jeong~San~Kim}\email{jkim@qis.ucalgary.ca}
\affiliation{
 Institute for Quantum Information Science,
 University of Calgary, Alberta T2N 1N4, Canada
}
\author{Taewan~Kim}\email{april02@snu.ac.kr}
\affiliation{
 Department of Mathematical Sciences,
 Seoul National University, Seoul 151-747, Korea
}
\author{Soojoon~Lee}\email{level@khu.ac.kr}
\affiliation{
 Department of Mathematics and Research Institute for Basic Sciences,
 Kyung Hee University, Seoul 130-701, Korea
}

\date{\today}

\begin{abstract}
We develop a three-party quantum secret sharing protocol
based on arbitrary dimensional quantum states.
In contrast to the previous quantum secret sharing protocols,
the sender can always control the state,
just using local operations,
for adjusting the correlation of measurement directions of three parties
and thus there is no waste of resource due to the discord between the directions.
Moreover, our protocol contains the hidden value
which enables the sender to leak no information of secret key to the dishonest receiver
until the last steps of the procedure.

\end{abstract}

\pacs{
03.67.Hk,  
03.67.Dd   
}
\maketitle

\section{Introduction}
In classical secret sharing \cite{Shamir,Blakley}, one party, say Alice,
wants to send her message to the other parties (Bob and Charlie) at
a distance. However Alice suspects that one of the others may be
dishonest, and she does not know who is the dishonest one. She
tries to divide the secret message into two pieces and give the proper relation
between them so that Bob and Charlie can decode the message
only if they cooperate in the same place.

Hillery {\it et al.}~\cite{HBB} first proposed a quantum secret sharing scheme
with a tripartite entangled state called the Greenberger-Horne-Zeilinger (GHZ) state~\cite{GHZ},
which was generalized into quantum secret sharing (QSS) protocols on any higher dimensional systems
using a $N$-party $N$-level singlet state of total spin zero~\cite{C}.
However, the protocols still have the restriction that the number of participants
should be the same as the dimension of each particle.

In this paper
we construct a QSS protocol which does not have such a limit,
and which contains a hidden value
controlling the correlation among outcomes of three parties.
Moreover, we show that
our protocol based on GHZ-like states can be more efficient than any previous QSS protocols,
by allowing Alice to manage and to rotate the states locally
according to the measurement directions of Bob and Charlie.

Most of quantum cryptographic protocols assure that
malicious eavesdropper cannot get the exact information about private key
and can be detected with a specific probability
when she measures a given state in the wrong direction.
Thus, in order for QSS protocols to be secure,
Eve's wrong measurement should give rise to uncertainty as much as possible.
On this account,
two mutually unbiased basis (MUB) measurements \cite{WF,Barnum,KR}
on $d$-dimensional quantum systems
play an important role in our protocol.

In Section~\ref{Sec:GHZ}
we consider the generalized Pauli operators acting on $d$-dimensional systems,
and derive MUBs and the GHZ-like states.
We provide our QSS protocol
based on the GHZ-like states in Section~\ref{Sec:Protocol},
and analyze the security of the protocol
for two cases of attacks in Section~\ref{Sec:Security},
where one is an eavesdropping by Eve,
and the other is the intercept-and-resend attack by dishonest person.
We finally summarize our results in Section~\ref{Sec:conclusion}.

\section{GHZ-like states on $d$-dimensional quantum systems}\label{Sec:GHZ}

In this section, we derive two MUBs and GHZ-like states on $d$-dimensional quantum systems,
and investigate their properties related with the security of our protocol.
First, we consider the generalized Pauli operators
acting on $d$-dimensional Hilbert space.

\begin{eqnarray}
\tilde{X}=\sum_{j=0}^{d-1}\ket{j+1}\bra{j},\quad
\tilde{Z}=\sum_{j=0}^{d-1}\omega^{j}\ket{j}\bra{j}, \\
\tilde{Y}=\tilde{X}\tilde{Z}=\sum_{j=0}^{d-1}\omega^{j}\ket{j+1}\bra{j},
\end{eqnarray}
where $\omega=e^{2\pi i/d}$ is a primitive $d$-th root of unity.
Let
\begin{equation}
\ket{k_x}=\frac{1}{\sqrt{d}}
\sum_{j=0}^{d-1}\omega^{-kj}\ket{j}.
\label{eq:k_x}
\end{equation}
Then $\ket{k_x}$ is an eigenstate of $\tilde{X}$ with eigenvalue
$\omega^{k}$.
Let
\begin{equation}
\ket{k_y}=
\begin{cases}
\displaystyle
\frac{1}{\sqrt{d}}
\sum_{j=0}^{d-1}\omega^{\frac{j^2-2kj-j}{2}}\ket{j} &
\hbox{if $d$ is odd,} \\
\displaystyle
\frac{1}{\sqrt{d}}
\sum_{j=0}^{d-1}\omega^{\frac{j^2-2kj-2j}{2}}\ket{j} &
\hbox{if $d$ is even.}
\end{cases}
\label{eq:k_y}
\end{equation}
Then $\ket{k_y}$ is an eigenstate of $\tilde{Y}$
with eigenvalue $\omega^{k}$ if $d$ is odd
and with eigenvalue $\omega^{k}\sqrt{w}$ if $d$ is even.

For each $d$, the set of eigenstates
$\{\ket{k_x}: k\in \Z_d\}$ of $\tilde{X}$ forms an orthonormal basis
for a $d$-dimensional quantum system,
and so does $\{\ket{k_y}: k\in \Z_d \}$ of $\tilde{Y}$.
Furthermore, they are mutually unbiased to each other, that is,
for any $k,~k' \in \Z_d$
$$|\inn{k_x}{k'_y}|=\frac{1}{\sqrt{d}.}$$
In our protocol, two MUB measurements,
$X=\{\ket{k_x}\bra{k_x}:k\in \Z_d\}$
and $Y=\{\ket{k_y}\bra{k_y}:k\in \Z_d\}$,
are alternatively used.

Let us construct a three-party entangled state
\begin{equation}
\ket{\Psi(\alpha)}_{XYY}= \frac{1}{d}\sum_{s+t+u=\alpha \pmod d}\ket{s_x}\ket{t_y}\ket{u_y},
\label{eq:psiXYY1}
\end{equation}
where $\alpha\in\Z_d$.
Then we can readily obtain
\begin{equation}
\ket{\Psi(\alpha)}_{XYY}=
\begin{cases}
\displaystyle
\dfrac{1}{\sqrt{d}}\sum_{j=0}^{d-1}\omega^{j(j-1-\alpha)}\ket{jjj} & \hbox{if $d$ is odd}, \\
\displaystyle
\dfrac{1}{\sqrt{d}}\sum_{j=0}^{d-1}\omega^{j(j-2-\alpha)}\ket{jjj} & \hbox{if $d$ is even}.
\end{cases}
\label{eq:psiXYY2}
\end{equation}
Similarly we can derive an entangled state $\ket{\Psi(\alpha)}_{XXX}$ as follows:
\begin{eqnarray}
\ket{\Psi(\alpha)}_{XXX}
&=&  \frac{1}{d}\sum_{s+t+u=\alpha \pmod d}\ket{s_x}\ket{t_x}\ket{u_x} \nonumber \\
&=&   \frac{1}{\sqrt{d}}\sum_{j=0}^{d-1}\omega^{-j \alpha}\ket{jjj}.
\label{eq:psiXXX}
\end{eqnarray}
It is easy to check that
$\ket{\Psi(1)}_{XYY}$ and $\ket{\Psi(0)}_{XXX}$ are the same for $d=2$,
and furthermore
both $\ket{\Psi(\alpha)}_{XYY}$ and $\ket{\Psi(\alpha)}_{XXX}$
are essentially equivalent to the standard $d$-dimensional GHZ state
up to local unitary operations.
In particular, it follows from Eqs.~(\ref{eq:psiXYY2}) and (\ref{eq:psiXXX}) that
\begin{eqnarray}
\ket{\Psi(\alpha)}_{XYY}
&=& (U\otimes I \otimes I)\ket{\Psi(\alpha)}_{XXX} \nonumber\\
&=&  \frac{1}{d}\sum_{s+t+u=\alpha \pmod d}U\ket{s_x}\ket{t_x}\ket{u_x},
\label{eq:XYY_XXX}
\end{eqnarray}
where
\begin{equation}
U=
\begin{cases}
\displaystyle\sum_{j=0}^{d-1}\omega^{j(j-1)}\ket{j}\bra{j} & \hbox{if $d$ is odd}, \\
\displaystyle\sum_{j=0}^{d-1}\omega^{j(j-2)}\ket{j}\bra{j} & \hbox{if $d$ is even}.
\end{cases}
\label{eq:U}
\end{equation}
In this point of view,
we call these states the {\em GHZ-like} states.

We now show that 
$\ket{\Psi(\alpha)}_{XYY}$ is the uniquely determined common eigenstate
of $XYY$, $YXY$ and $YYX$
with respect to eigenvalue $\omega^{\alpha}$ if $d$ is odd
($\omega^{\alpha+1}$ if $d$ is even).
Let $d$ be odd
and assume that an arbitrary 3-qudit pure state
$\ket{\phi}=\sum_{j,k,l}a_{jkl}\ket{jkl}$ satisfies
\begin{equation}
XYY\ket{\phi} = YXY\ket{\phi}=YYX\ket{\phi} =\omega^{\alpha}\ket{\phi}.
\end{equation}
It follows from straightforward calculations that
\begin{equation}
\ket{\phi}= \frac{1}{\sqrt{d}} \sum_{j=0}^{d-1} \omega^{j(j-1-\alpha)}\ket{jjj}
=\ket{\Psi(\alpha)}_{XYY}.
\end{equation}
Similarly, if $d$ is even, we also have
\begin{equation}
\ket{\phi}=\frac{1}{\sqrt{d}} \sum_{j=0}^{d-1}\omega^{j(j-2-\alpha)}\ket{jjj}
=\ket{\Psi(\alpha)}_{XYY}.
\end{equation}
Moreover, we can see that
$\ket{\Psi(\alpha)}_{XYY}=\ket{\Psi(\alpha)}_{YXY}=\ket{\Psi(\alpha)}_{YYX}$.
Hence, if Alice, Bob and Charlie measure $\ket{\Psi(\alpha)}_{XYY}$
by $XYY$, $YXY$, or $YYX$,
then they obtain outcomes $s$, $t$ and $u$ satisfying that $s+t+u=\alpha \pmod d$, respectively.

\section{Our Protocol}\label{Sec:Protocol}

In QSS, Bob and Charlie obtain the Alice's private key from the correlation of outcomes,
given by measuring a three-party entangled quantum systems.
In fact, Bob and Charlie can get Alice's information
by the joint measurement such as Bell measurement
if they are together at same place.
This is the same situation as QKD like BB84 or EPR protocols~\cite{BB,E}.

However, QSS protocol proceeds in the condition that they are far away from each other
and measure their states locally. Non-locality and entanglement distributed between them are,
after all, used to give a correlation between their classical outcomes by local measurements.
Therefore, one of the most important problem in QSS is
how Alice sends an entangled state to Bob and Charlie
securely against eavesdropping by any exterior Eve
and the intercept-and-resend attack
by an interior dishonest person.
In order to construct the QSS protocol satisfying the above conditions,
we use two MUB measurements given in Section~\ref{Sec:GHZ}.

\begin{enumerate}

\item Alice prepares a GHZ-like state,
$\ket{\Psi(\alpha)}_{XYY}$,
and sends Bob and Charlie the last two particles, respectively.
Alice repeats this step $2n$ times,
and all participants store their particles in the order received.

\item Bob and Charlie publicly announce
the fact that they have already received all $2n$ particles from Alice,
and then they measure their own qudits
after deciding one of measurement directions $X$ and $Y$ randomly.

\item\label{step:test0}
Alice informs Bob and Charlie a randomly chosen $2n$ bit string $\mathbf{b}$,
each entry of which is either $0$ or $1$.
Then for $i$-th particles corresponding to $\mathbf{b}_i=1$
Alice requires Bob and Charlie to announce their measurement outcomes
and directions in the order randomly determined as either
[(i) Bob's outcome, (ii) Charlie's outcome, (iii) Charlie's direction, (iv) Bob's direction]
or [(i) Charlie's outcome, (ii) Bob's outcome, (iii) Bob's direction, (iv) Charlie's direction].

\item\label{step:test} Alice properly measures her $i$-th particles corresponding to $\mathbf{b}_i=1$
in the direction correlated with measurement of Bob and Charlie
as in TABLE~\ref{table}.

\begin{table}
\begin{center}
\begin{tabular}{c|c|c|c}
  \hline
  \hline
  State & Bob & Charlie & Alice \\
  \hline
  $\ket{\Psi(\alpha)}_{XYY}$ & $Y$ & $Y$ & $X$ \\
  $\ket{\Psi(\alpha)}_{XYY}$ & $Y$ & $X$ & $Y$ \\
  $\ket{\Psi(\alpha)}_{XYY}$ & $X$ & $Y$ & $Y$ \\
  $\ket{\Psi(\alpha)}_{XYY}$ & $X$ & $X$ & $UXU^{\dagger}$ \\
  \hline
  \hline
\end{tabular}
  \caption{
Alice's measurements corresponding to Bob's and Charlie's:
$U$ is the local unitary operation which transforms $\ket{\Psi(\alpha)}_{XXX}$
into $\ket{\Psi(\alpha)}_{XYY}$ in Eq.~(\ref{eq:U}).}\label{table}
\end{center}
\end{table}

\item If Alice finds any error from all participants' measurement outcomes
in Step~\ref{step:test}, then she aborts the protocol.
Otherwise,
they discard the particles for the test, and
Alice lets Bob and Charlie announce
their measurement directions for the particles left after the test.

\item Alice properly measures her particles
in the direction perfectly correlated with measurement of Bob and Charlie
as in TABLE~\ref{table}.

\item When Bob and Charlie collaborate to obtain Alice's information,
Alice announces the hidden value $\alpha$ to Bob and Charlie.
Then they can derive her private key string from the outcome correlation,
$s+t+u=\alpha \pmod d$.
\end{enumerate}

Note that it is possible to use a string
consisting of different hidden values for GHZ-like states,
instead of the fixed $\alpha$.

\section{Security}\label{Sec:Security}

\subsection{Eavesdropping by exterior Eve}

In section~\ref{Sec:GHZ},
we have shown that $\ket{\Psi(\alpha)}_{XYY}$ is the unique pure three-party quantum state
invariant under operators $XYY$, $YXY$ and $YYX$ simultaneously,
with respect to an eigenvalue $\omega^{\alpha}$ if $d$ is odd
($\omega^{\alpha+1}$ if $d$ is even).

This means that if
\begin{equation}
\ket{\Psi}=\sum_{j,k,l = 0}^{d-1}a_{jkl}\ket{jkl}_{ABC} \ket{R_{jkl}}_E
\label{eq:generalPsi}
\end{equation}
successfully passes the test of our protocol
then $\ket{\Psi}$
should be a product state
\begin{equation}
\ket{\Psi}=\ket{\Psi(\alpha)}_{XYY}\otimes\ket{R}_E.
\label{eq:productPsi}
\end{equation}
In other words, after the test of our protocol,
Eve is perfectly separated and
the perfect correlation, $s+t+u=\alpha \pmod d$, is securely preserved among all participants.
Therefore, our protocol is secure against any exterior Eve's eavesdropping.

\subsection{Intercept-and-resend attack by interior dishonest party }
In this section, we consider the case
that one of receivers Bob and Charlie changes his mind
and tries to obtain Alice's private key alone.
Suppose a dishonest person (Bob) performs the intercept-and-resend attack on Charlie's particles.

First, Bob can intercept, measure by predicting the measurement direction of Charlie,
and resend the collapsed state to him.
If Bob and Charlie measure Charlie's original states in the same directions,
then Bob can obtain the information about Alice's private key alone
after knowing the hidden value $\alpha$.
Although Bob performs measurements in the directions different from Charlie,
his attacks can be unexposed with probability $1/d$.
Therefore, the exposed probability is not less than $1-\left(\frac{d+1}{2d}\right)^n$
during the test procedure
and we can find out that
the higher dimensional system provides us with the better security for QSS protocol.
This is due to the fact that the number of eigenspaces of measurement linearly increases
as the dimension of system gets higher,
and that it is also difficult for Bob to obtain the same result as Charlie's
when $n$ is sufficiently large.

We now assume that Bob possesses all states Alice sent
and gives Charlie one sides of $d$-dimensional bipartite (maximally entangled) states.
In Step~\ref{step:test0} of our protocol,
the measurement directions and outcomes of Bob and Charlie
are alternately announced in a specific way.
As in \cite{KKI}, this procedure prevent dishonest Bob
from cheating the other members.
Therefore,
our protocol is also secure against
intercept-and-resend attacks by an interior dishonest member.

\section{Conclusions}\label{Sec:conclusion}
We have presented a 3-party $d$-level QSS protocol.
To construct a QSS protocol on arbitrary $d$-dimensional quantum systems,
we have derived MUBs on Hilbert space $\mathbb{C}^d \otimes \mathbb{C}^d \otimes \mathbb{C}^d$,
which guarantees the security of our protocol.
Especially,
with the explicit formula for the exposed probability,
we have shown that the higher dimensional system assures
the better security for QSS protocol.

In addition to the security,
our protocol is more efficient than any other protocols
since the number of discarding entangled states is minimized in our protocol
by controlling Alice's measurements according to measurements of Bob and Charlie.
Furthermore, in contrast to the previously known QSS protocols,
Bob and Charlie have no information about Alice's private key
because of the hidden value or string $\alpha$,
although he is not detected in the middle of test.

\section*{Acknowledgments}
D.P.C. was supported by the Korea Science and Engineering Foundation
(KOSEF) grant funded by the Korea government (MOST) (No.~R01-2006-000-10698-0), and
S.L. was supported by the Korea Research Foundation Grant funded by the Korean Government
(MOEHRD, Basic Research Promotion Fund) (KRF-2007-331-C00049).

\end{document}